\documentclass[prc,aps,superscriptaddress,showpacs,twocolumn]{revtex4-1}
\usepackage{bm}
\usepackage{dcolumn}
\usepackage{graphicx}
\usepackage{amsmath}
\usepackage{amssymb}
\usepackage{color}

\newcommand{\be}{\begin{equation}}
\newcommand{\ee}{\end{equation}}

\newcommand{\beqa}{\begin{eqnarray}}
\newcommand{\eeqa}{\end{eqnarray}}

\begin{document}
\title{
{The nuclear electric polarizability of $^6$He}}

\author{R.\ Goerke}
\email[E-mail:~]{rgoerke@physics.utoronto.ca}
\affiliation{Department of Physics, University of Toronto,
60 St. George St., Toronto, Ontario, M5S 1A7, Canada}
\affiliation{TRIUMF, 4004 Wesbrook Mall, Vancouver, BC, V6T 2A3, Canada}
\author{S.\ Bacca}
\email[E-mail:~]{bacca@triumf.ca}
\affiliation{TRIUMF, 4004 Wesbrook Mall, Vancouver, BC, V6T 2A3, Canada}
\author{N.\ Barnea}
\email[E-mail:~]{nir@phys.huji.ac.il}
\affiliation{Racah Institute of Physics, Hebrew University, 91904, Jerusalem,
Israel}

\date{\today}

\begin{abstract}
We present an estimate of the nuclear electric polarizability of 
 the $^6$He halo nucleus based on  six-body microscopic calculations.
Wave functions are obtained from  semi-realistic two-body interactions using
the hyperspherical harmonics expansion method.  The polarizability is calculated as a
sum rule of the dipole response function using the Lanczos algorithm and also
by integrating the photo-absorption cross section calculated via the Lorentz
integral transform method.  We obtain  
$\alpha_E=1.00(14)$ fm$^3$, which is much smaller than the published value
$\alpha_E^{\rm exp}=1.99(40)$ fm$^3$~\cite{Moro} extracted from experimental
data. This points towards a potential disagreement between  microscopic
theories and experimental observations. 
\end{abstract}

\maketitle

\section{Introduction}
The nuclear electric polarizability $\alpha_E$ is related to the response of a nucleus to an externally applied electric field. It is an interesting observable because it encapsulates information
about the excitation spectrum of a nucleus. Recently, it has attracted a lot of
attention both for light nuclei, see {\it e.g.}~\cite{Moro}, and for heavy
nuclei, see {\it e.g.}~\cite{Pb_RCNP}. 
For light systems  the nuclear polarizability is relevant
in the extraction of nuclear quantities from atomic spectroscopic
measurements. The atomic energy levels are affected by  polarization of the
nucleus due to the electric field of the surrounding electrons. Such nuclear
structure correction, which is proportional to  
$Z^3\alpha_E/a_0$ \cite{noert} where $a_0$ is the Bohr radius, needs to be
considered in the sophisticated quantum electrodynamics  calculations of the
atomic levels that  allow the extraction of charge radii from isotope shifts
measurements of unstable nuclei, see Refs.~\cite{Wan04, Maxime} and
\cite{Mue07} for $^6$He and $^8$He, respectively. 
An even larger effect of the nuclear structure correction coming from the polarizability is expected in muonic atoms, as the  muon mass 
is larger than the electron mass and the orbital radius is smaller. 
This will be relevant to the proposed $\mu ^4$He and  $\mu ^3$He experiments \cite{rolf_exp} that aim at measuring the nuclear charge radius of $^4$He and $^3$He from the Lamb-shift, to be compared to electron scattering data. 

 The nuclear electric polarizability of Helium isotopes is interesting for the several above mentioned reasons.
It has been already directly measured or extracted from experimental data for
the $^{3,4,6}$He isotopes \cite{Moro}. In the case of $^3$He it is worth
mentioning that the data from elastic scattering on Pb at energies below the
Coulomb barrier  \cite{Goe91} are in disagreement with estimates based on
calculations of the photo-absorption cross section~\cite{Moro}, the latter
being about a factor of 2 smaller. It is also worth noticing that the
theoretical calculations are in agreement with photo-absorption 
experiments, and that the band
spanned by using different Hamiltonians in the calculations is smaller than
the difference between the data taken from photo-absorption cross section and from
ion scattering experiments. The data analysis involved in the  latter approach
is quite delicate, because one has to separate effects of the nuclear force
from Coulomb effects.

In Ref.~\cite{Stetcu} the polarizabilities of several Hydrogen and Helium isotopes were calculated with  {\it ab-initio} methods.
 Among the Helium isotopes, $^3$He and $^4$He were dealt with, but no prediction for $^6$He was provided. It is the aim of this paper to fill this gap.

The $^6$He is known as a  halo nucleus, where a tightly bound  $^4$He core is surrounded
by two neutrons~\cite{Tan96}.  It happens to be the lightest of the known halo nuclei
 and it is a Borromean nucleus, because the
two-neutron and neutron-core subsystems are unbound, but the
three-body system is held together. 
Due to the very small separation energy which characterizes halo nuclei, one expects the
polarizability of $^6$He to be much larger than that of the tightly bound $^4$He isotope.
Experimental data indicate this behavior.
In this paper we would like to see whether microscopic calculations
reproduce the experimental values and lead to a result where 
 $\alpha_E(^6{\rm He}) \gg \alpha(^4{\rm He}$).

We perform a microscopic study of the nuclear polarizability $\alpha_E$ for
$^6$He and compare it to $^4$He. We limit our study  to simple semi-realistic
two-body forces.  
For that we use the  hyperspherical harmonics method with an effective interaction,
EIHH, to speed up the convergence~\cite{Nir,EIHH}. 
 The polarizability is calculated as a sum rule of the dipole response function using the Lanczos algorithm and also integrating the photo-absorption cross section calculated with the Lorentz integral transform method~\cite{ELO94}.

This paper is organized as follows. In Section II we describe in details the theoretical calculation of the polarizability.
In Section III we present our results and in Section IV we make a comparison 
with experiment. Finally, we conclude in Section V.

\section{Theoretical aspects}

The nuclear electric polarizability in the unretarded dipole approximation
is defined by the expression
\begin{equation}
\alpha_E=2 \alpha \sum_{n\ne0} \frac{|\langle n| {D_z}| 0\rangle|^2}{E_n -E_0}\,
\label{pol}
\end{equation}
where $\alpha$ is the fine structure constant, $D_z$ is the unretarded dipole
operator and  $E_{0/n}$ are the energies of the nuclear ground and excited states $|0\rangle$ and $|n\rangle$, respectively.
This observable is clearly related to the photo-absorption cross section and to the dipole
response function.
The photo-absorption cross section $\sigma_\gamma(\omega)$ of a nucleus 
is given by 
\begin{equation}
\sigma_\gamma(\omega)= 4\pi^2 \alpha
\, \omega\, R(\omega)\,,\label{sigma} 
\end{equation}
where $R(\omega)$ is the response function.
In the unretarded dipole approximation 
\begin{equation}
R(\omega) = \sum_{n,\bar{0}} |\langle n| D_z| 0\rangle|^2\,
            \delta(\omega-E_n+E_0) \,,
\end{equation}
where $\bar{0}$ indicates an average on the initial angular
momentum projections.
The dipole operator is given by $D_z=\sum_{i=1}^A z_i \tau_i^{3}/2$,
where $A$ is the number of nucleons and $\tau^3_i$ and $z_i$ are the third
component of the isospin  
operator and the coordinate of the $i$th particle in the center of mass frame, respectively.
One can recover the expression for $\alpha_E$ in Eq.~(\ref{pol}) by calculating 
sum rules of the photo-nuclear cross section.
The various  moments of $\sigma_{\gamma}$ are defined as 
\begin{equation}
m_n(\bar\omega) \,\,\, \equiv \int_{\omega_{th}}^{\bar\omega}\,d\omega \,\omega^n\,
\sigma_\gamma(\omega)\,,\label{moments} 
\end{equation}
where $\omega$ is the photon energy and $\omega_{th}$ and $\bar\omega$ indicate threshold energy and upper integration limit, respectively.  
Assuming that $\sigma_\gamma(\omega)$ converges to zero 
and utilizing the closure  of the  eigenstates of 
the nuclear Hamiltonian $H$, one can relate the polarizability to the
  $n=-2$ sum rule, 
\begin{eqnarray}
\alpha_E&=& \frac{m_{-2}(\infty)}{2\pi^2}
 = 2 \alpha
\,\sum_n \frac{|\langle n| D_z|0\rangle|^2}{E_n-E_0}\,.\label{mPSR}
\label{alphaEPSR}
\end{eqnarray}
The polarizability $\alpha_E$ can be calculated
with the Lanczos algorithm using a proper pivot. It is useful to rewrite Eq.~(\ref{alphaEPSR}) 
as 
\begin{eqnarray}
\nonumber
\alpha_E&=& 2\alpha \langle 0| {D}^{\dagger}_z \frac{1}{H-E_0} D_z |0\rangle\,\\
\label{DHm1D}
&=& - 2\alpha \langle 0|D_z^{\dagger}D_z|0\rangle \langle \phi_0| \frac{1}{E_0-H} |\phi_0\rangle\,,
\end{eqnarray} 
with
\begin{equation} \label{start}
  |\phi_0\rangle =\frac{D_z|0\rangle }{\sqrt{\langle
  0| D_z D_z|0 \rangle }} \,.
\end{equation}
Starting from the ``pivot'' of Eq.~(\ref{start})
where the ground state $|0\rangle$ is obtained by solving the Schr\"{o}dinger equation, 
 $\alpha_E$ can be expressed as a continued fraction containing the Lanczos coefficients~\cite{Dagotto}
\begin{equation}
   a_i=\langle \phi_i |H|\phi_i \rangle \mbox{,}\,\,\,\,\,\,
b_i=\langle \phi_{i+1}|H|\phi_i \rangle \mbox{,}
\end{equation}
where the $|\phi_i\rangle$ form the Lanczos orthonormal basis $\{|\phi_i
\rangle \mbox{,}i=0,\ldots \}$. In fact one has
\begin{equation}
\alpha_E= - 2\alpha \langle 0|D_z^{\dagger}D_z|0\rangle \cfrac{1}{E_0-a_0- \cfrac{b_1^2}{E_0-a_1 -\cfrac{b^2_2}{E_0-a_3 \dots}}} \,.
\label{cfrac}
\end{equation}

In this work we calculate the polarizability in two different ways. On the one hand we
utilize Eq.~(\ref{cfrac}). 
On the other hand we obtain $m_{-2}$ by integrating our results for the 
 total photo-absorption cross section  calculated with
the Lorentz integral transform (LIT) method \cite{ELO94}.
In Ref.~\cite{He6Sonia1, He6Sonia2} 
we have presented microscopic
calculations of the $^6$He $\sigma_{\gamma}$ with semi-realistic potential
models. Here we  use lager model spaces  which are nowadays available. 
The LIT, an integral transform with a Lorentzian kernel, is defined as 
\begin{equation}
{\cal L}(\sigma_R,\sigma_I) = \int d\omega {\frac {R(\omega)} {(\omega-\sigma_R)^2 
                              + \sigma_I^2} } \,.
\end{equation}
The LIT is also typically calculated   using the Lanczos technique explained above, (see~\cite{Mario}). In fact it can be re-expressed as
\begin{equation} \label{loreins}
{\cal  L}(\sigma_R, \sigma_I)=-\frac{1}{\sigma_I}\,
            \langle 0 | D_z D_z|0\rangle\,
            \mbox{Im} \,\{\langle \phi_0 |
                      \frac{1}{z-H}|\phi_0\rangle\} \mbox{} 
\end{equation}
with $z=E_0+\sigma_R+ i\sigma_I$. 
It is evident that the LIT in (\ref{loreins})
is also a 
 continued fraction as in Eq.~(\ref{cfrac}), where $E_0$ is replaced by a complex  $z=E_0+\sigma_R+ i\sigma_I$.
Once  ${\cal L}(\sigma_R,\sigma_I)$ is calculated, one can invert the LIT \cite{inversion}  to get 
$R(\omega)$, and thus $m_{-2}$.
The two methods have to agree within the numerical uncertainty. However, with
the first method one avoids the complications
introduced by the inversion procedure. 

Given the Hamiltonian $H$, the calculation of $\alpha_E$ in both ways is based on the EIHH~\cite{EIHH}
expansion of the wave function.
This  approach  is translational invariant, being constructed with the Jacobi coordinates.
We use different semi-realistic potential models for our calculations. Following 
Ref.~\cite{He6Sonia1}, we will use the
 Minnesota (MN) potential \cite{Minnesota} 
\begin{eqnarray}
\nonumber
V_{ij}&=&\left[V_R+\frac{1}{2}\left(1+P^{\sigma}_{ij}\right)V_T+\frac{1}{2}\left(1-P^{\sigma}_{ij}\right)V_S\right] \times \\
\label{mnu}
&&\left[\frac{1}{2}u+\frac{1}{2}\left(2-u\right)P^r_{ij}\right] \,, 
\end{eqnarray}
where $P_{ij}^{\sigma,r}$ are spin and space-exchange
operators, $V_{R,T,S}$ are parameterized as linear combination of Gaussians of
the two-body relative distance and $u$ is a parameter. 
This force reproduces the $S-$wave nucleon-nucleon phase shifts and correctly
binds the deuteron. It renormalizes effects of the tensor force into its
central component. 
A typical value for $u$ in the Minnesota potential is $u=1$, as we used in~\cite{He6Sonia1}. Here we will explore the variation of this parameter by choosing $u\ge1$.
The mixing parameter $u$ does not affect the dominant $^1S_0$ and $^3S_1$
waves in the NN interaction, but only affects the $s=1$, $t=1$ channels, where
the dominant components are the $P-$waves ($^1P_1$ and $^3P_{0,1,2}$).  
For $u=1$ there are no $P-$waves, they contribute only for $u>1$.
Thus, changing $u$ mostly affects $^6$He, without substantially changing $^4$He.
 Because in~\cite{He6Sonia1, He6Sonia2} we also used the Malfliet-Tjon (MTI-III) \cite{MTI-III},
 and the Argonne AV4' \cite{AV4'} potentials, we will present some results  with these interactions as well.
The Minnesota potential has been recently used in  a microscopic cluster model  calculation of $^6$He  \cite{Brida}  and in the Gamow shell model approach \cite{Papadim}  for 
$^6$He and $^8$He.

\section{Results and Discussion}

\begin{figure}
\includegraphics[scale=0.35,clip=]{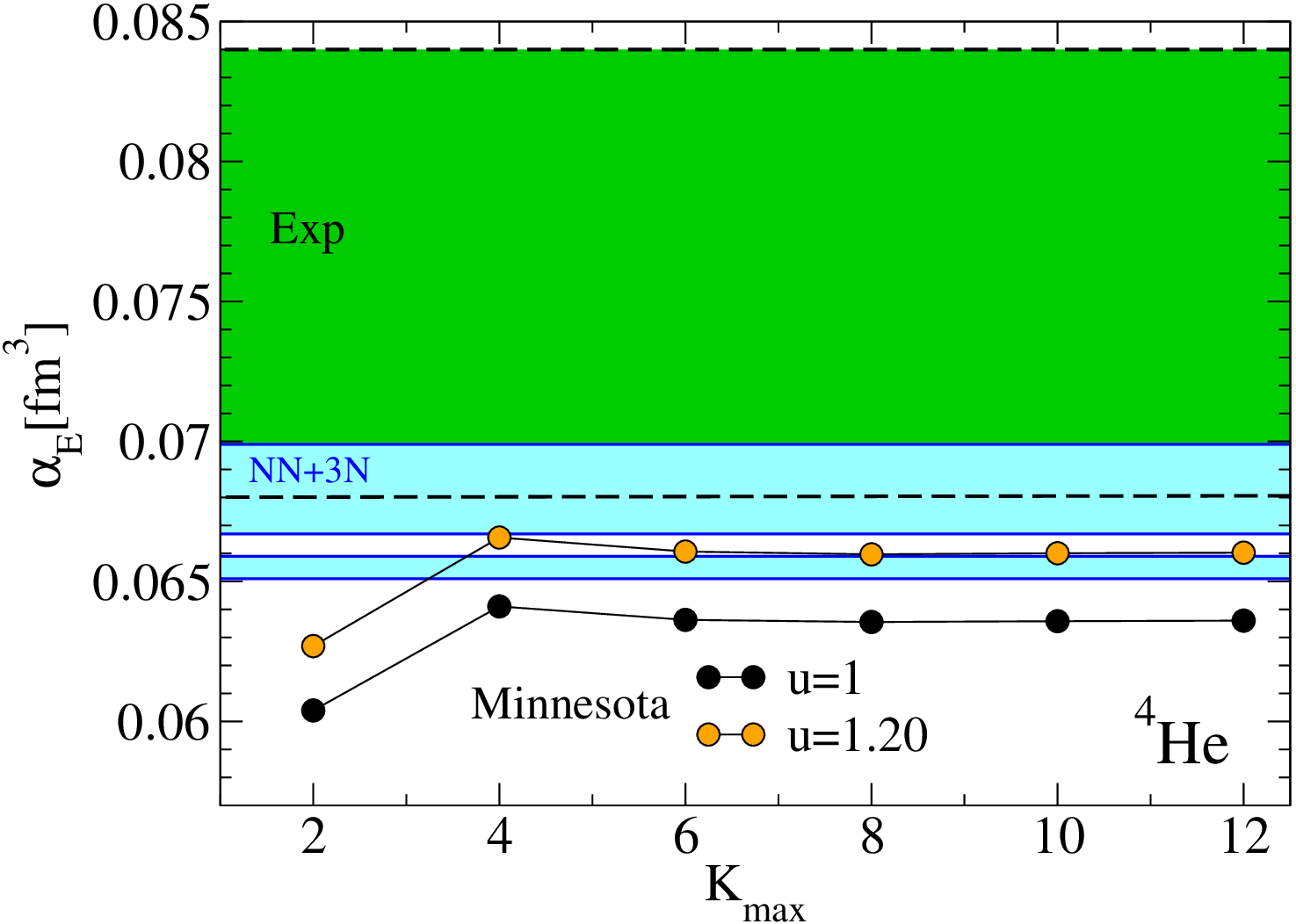}
\caption{(Color online) $^4$He polarizability:  calculations with the Minnesota potential for two different $u$ values are shown as a function of the grand-angular momentum quantum number $K_{\rm max}$.
The polarizabilities obtained from realistic two- and three-body interactions \cite{Stetcu, Gazit} are  presented as a light (blue) band. Experimental data from \cite{Moro, Friar} are given by the dark (green) band.
} 
\label{fig_He4}
\end{figure}

The main focus of this work is to study the $^6$He polarizability. 
We start, however, the discussion with the $^4$He nucleus.
In Fig.~\ref{fig_He4}, we show the results of $\alpha_E$
calculated via the Lanczos coefficients, as in Eq.~(\ref{cfrac}). 
The ground state $|0\rangle$ and the Lanczos ``pivot'' $ |\phi_0 \rangle$
are given in terms of the EIHH expansions.  While for the ground state the expansion is 
characterized by an even hyperspherical grand-angular quantum number $K_{\rm max}$ and total 
isospin $T=0$, $T_z=0$, $ D_z|0 \rangle$ has to be expanded on odd grand-angular quantum number $K'_{\rm max}$, where the isospin in the final state is $T'=1$.  Fig.~\ref{fig_He4} shows the convergence of $\alpha_E$ as a function of $K_{\rm max}$, where for each point $K_{\rm max}+1$ is used for the Lanczos ``pivot''.  We show our results for the Minnesota potential with $u=$1 and 1.20.
The convergence is very good, the dependence on $u$ is mild and the results are
very close to calculations where realistic nucleon-nucleon (NN) and three-nucleon (3N) forces have been used. For the latter one, results for effective field theory potentials were presented in \cite{Stetcu} leading to $\alpha_E=0.0683(8)(14)$ fm$^3$ (corresponding to the upper light (blue) band in Fig.~\ref{fig_He4}).
The error bar of this calculation is accounting for the convergence error $0.0008$ fm$^3$ 
 and also for the uncertainty in the underlying dynamics $0.0014$ fm$^3$.
We also show the  results  of   $\alpha_E$ for the Argonne $v_{18}$ two-body force and Urbana IX three-body force of Ref.~\cite{Gazit} leading to $\alpha_E=0.0655(4)$
 (corresponding to the lower light (blue) band in Fig.~\ref{fig_He4}), where the error bar comes from convergence only.   
The experimental data are shown as a darker (green) band. These include the more recent evaluation of Ref.~\cite{Moro} based on the Arkatov {\it et al.}~\cite{Arkatov} experimental measurement of the photo-absorption cross section and an older result reported in Ref.~\cite{Friar}, based on earlier measurements by Arkatov {\it et al.}~\cite{Arkatov_old}. We would like to point out that the semi-realistic Minnesota potentials lead to a value of the polarizability which is consistent with realistic calculations and is only about 15$\%$ smaller than the average value in the experimental band.

We can also calculate the polarizability by integrating the photoabsorption cross section obtained with the LIT method. We get perfect agreement as with the Lanczos coefficients. For example, for the standard Minnesota potential where $u=1$ and for a $K_{\rm max}=12/13$ model space, the Lanczos method gives 
$\alpha_E=0.06360$ fm$^3$ and integrating $\sigma_{\gamma}$ up to 120 MeV we get 0.06336 fm$^3$, with just a $0.4\%$ difference.

\begin{figure}
\includegraphics[scale=0.35,clip=]{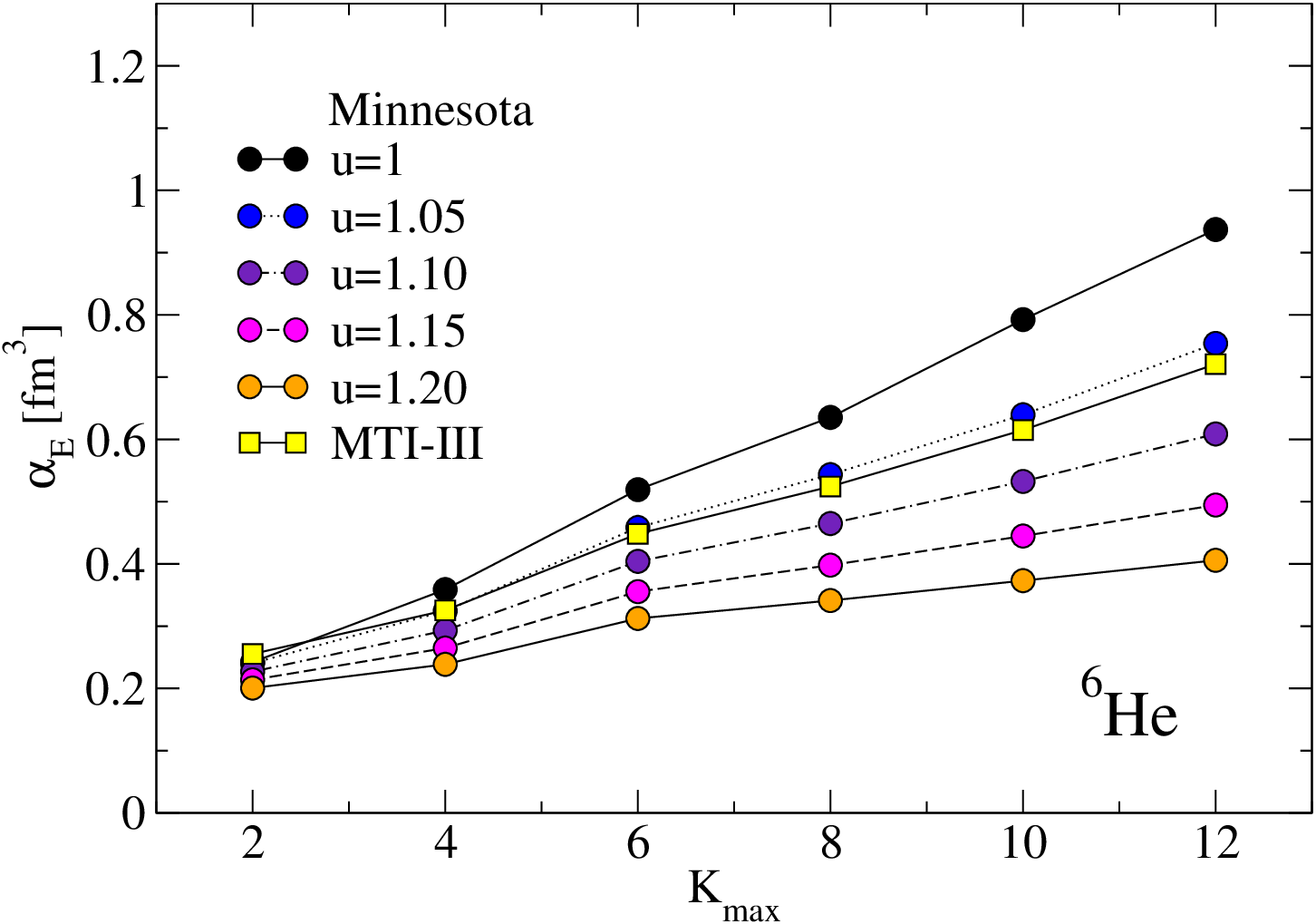}
\caption{(Color online) The polarizability of $^6$He 
as a function of the grand-angular momentum $K_{\rm max}$ for different semi-realistic interactions: the Minnesota potential with $u=1-1.2$ and the MTI-III potential.} 
\label{fig_alphaE}
\end{figure}

Now we move to the $^6$He nucleus.
We first  calculate $\alpha_E$ from the Lanczos coefficients.
Also in this case the ground state is  expanded on even hyperspherical grand-angular quantum number $K_{\rm max}$, but the total 
isospin  is $T=1$, $T_z=-1$, and $ D_z|0 \rangle$  is expanded on odd $K_{\rm max}'=K_{\rm max}+1$. In this case though, the final isospin can be
$T'=1$ or $T'=2$. This leads to two possible isospin channels that are calculated
separately and that open up at different energies. Experimentally, the $T=1$ channel opens up at photon energy $\omega_{th}=0.975$ MeV, while the $T=2$ channel opens up at $\omega_{th}=22.77$ MeV,  with $\gamma~^6$He $\rightarrow$ $^3$H$\,n\,n\,p$.
 Due to the inverse energy weight in  Eq.~(\ref{mPSR}), 
the $T=2$ channel is expected to be less relevant to $\alpha_E$.
From our calculations we find that the percentage contribution of the T=2 isospin channel to the total polarizability changes from $2\%$ to $4\%$ when varying $u$ from 1 to 1.20 in the Minnesota potential.    

\begin{table}
\caption{Results of the EIHH calculation with $K_{max}=12$ for the different $u$ value of the Minnesota potential. The values for the energies are in MeV.} 
\label{Table1}
\begin{center}
\begin{tabular}
{c | lll} 
\hline
Potential & $E_0$($^4$He) & $E_0$($^6$He) & $S_{2n}$($^6$He)\\
\hline\hline
Minnesota &  &  & \\
u=1.00 & -29.949 & -30.45 & 0.50\\
u=1.05 & -29.978 & -31.13 & 1.15\\
u=1.10 & -30.007 & -31.88 & 1.87\\
u=1.15 & -30.037 & -32.72 & 2.68\\
u=1.20 & -30.069 & -33.65 & 3.59\\
MTI-III & -30.760 & -32.24  & 1.48\\
\hline\hline
\end{tabular}
\end{center}
\end{table}

In Fig.~\ref{fig_alphaE}, we present a similar plot as in Fig.~\ref{fig_He4} for $^6$He with semi-realistic interactions.
 We observe a much slower convergence of $\alpha_E$ for $^6$He than for $^4$He with all the potentials employed.  
By looking at the different $u$ values in the Minnesota potential, we see that
the convergence rate and the value of $\alpha_E$  substantially change  with $u$.
This is related to the variation of the binding energy, and consequently of the two-neutron separation energy, whose  numerical values are shown in Table~\ref{Table1} for completeness.
By increasing $u$ we are adding more $P-$wave interactions, which bring
additional binding to the $^6$He nucleus, while leaving $^4$He almost
unaffected. Naively, this makes $^6$He more tightly bound and thus more
difficult to polarize,  i.e., $\alpha_E$ gets smaller. For the value of $u=1$
the convergence of the polarizability is particularly slow, due to the fact
that $^6$He is barely bound, with  $S_{2n}=0.56$ MeV which is about a factor
of 2 smaller than the experimental value. 
With the MTI-III potential we get a convergence pattern which is close to the Minnesota potential for $u=1.05$, because the prediction of $S_{2n}$ is similar with these two potentials, see Table~\ref{Table1}. The information that one gains from Fig.~\ref{fig_alphaE} is that by increasing $S_{2n}$ we can change the overall slope of the convergence pattern of $\alpha_E$.

\begin{figure}
\includegraphics[scale=0.35,clip=]{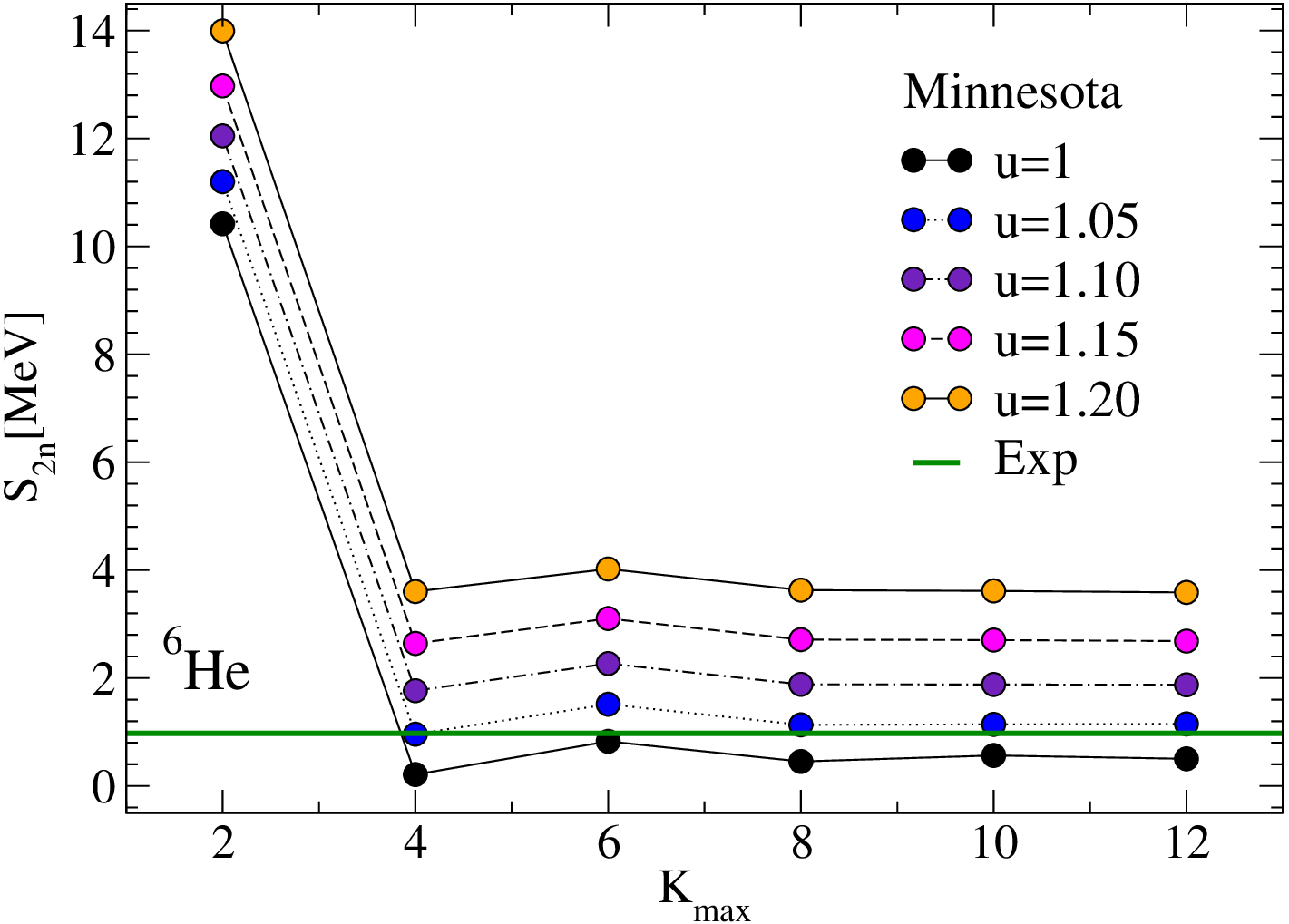}
\caption{(Color online) $^6$He two-neutron separation energy as a
function of the grand-angular momentum $K_{\rm max}$ for the Minnesota potential and different $u$ parameters. The experimental value
is also shown. 
} 
\label{fig_S2n}
\end{figure}
For any considered value of $u$ though, it is clear that
our calculations reproduce the fact that the polarizability of the halo nucleus of $^6$He is much larger than that of the tightly bound $^4$He, the ratio being almost an order of magnitude.

Here we would like to point out that Brida and Nunes
in \cite{Brida}  have used the Minnesota potential with $u=1.15$ in a microscopic cluster model 
and obtained a separation energy $S_{2n}=0.90(5)$ MeV. 
This result is  different from the value we obtain and report
in Table~\ref{Table1}. Their calculation is performed without the Coulomb force, but its effect cancels in the separation energy. Because for $^4$He the value reported in \cite{Brida} for the binding energy is -30.85 MeV, which is in agreement with our value of -30.86(1) (with no Coulomb force), we think that the difference is due to the cluster assumption  made for $^6$He. We do not make such assumption and in convergence, the EIHH result is exact. 
In  Fig.~\ref{fig_S2n}, we show that
the separation energy $S_{2n}$ is very well converged within the model space available for all
these potentials.

We can also calculate the polarizability by integrating the photoabsorption cross section obtained with the LIT method and compare it to the above results. We quote numbers for the  Minnesota potential with $u=1.05$ in the largest available model space $K_{\rm max}=12/13$.
 The Lanczos method gives 
$\alpha_E=0.7542$ fm$^3$ and integrating $\sigma_{\gamma}$ up to 40 MeV (60 MeV) we get 0.7711 fm$^3$ (0.7827) fm$^3$. 
Integrating the cross section we have a 3-4$\%$ difference, which is due
  to the fact that LIT is not completely converged and the inversion procedure
  introduces some numerical error.

\begin{figure}
\includegraphics[scale=0.35,clip=]{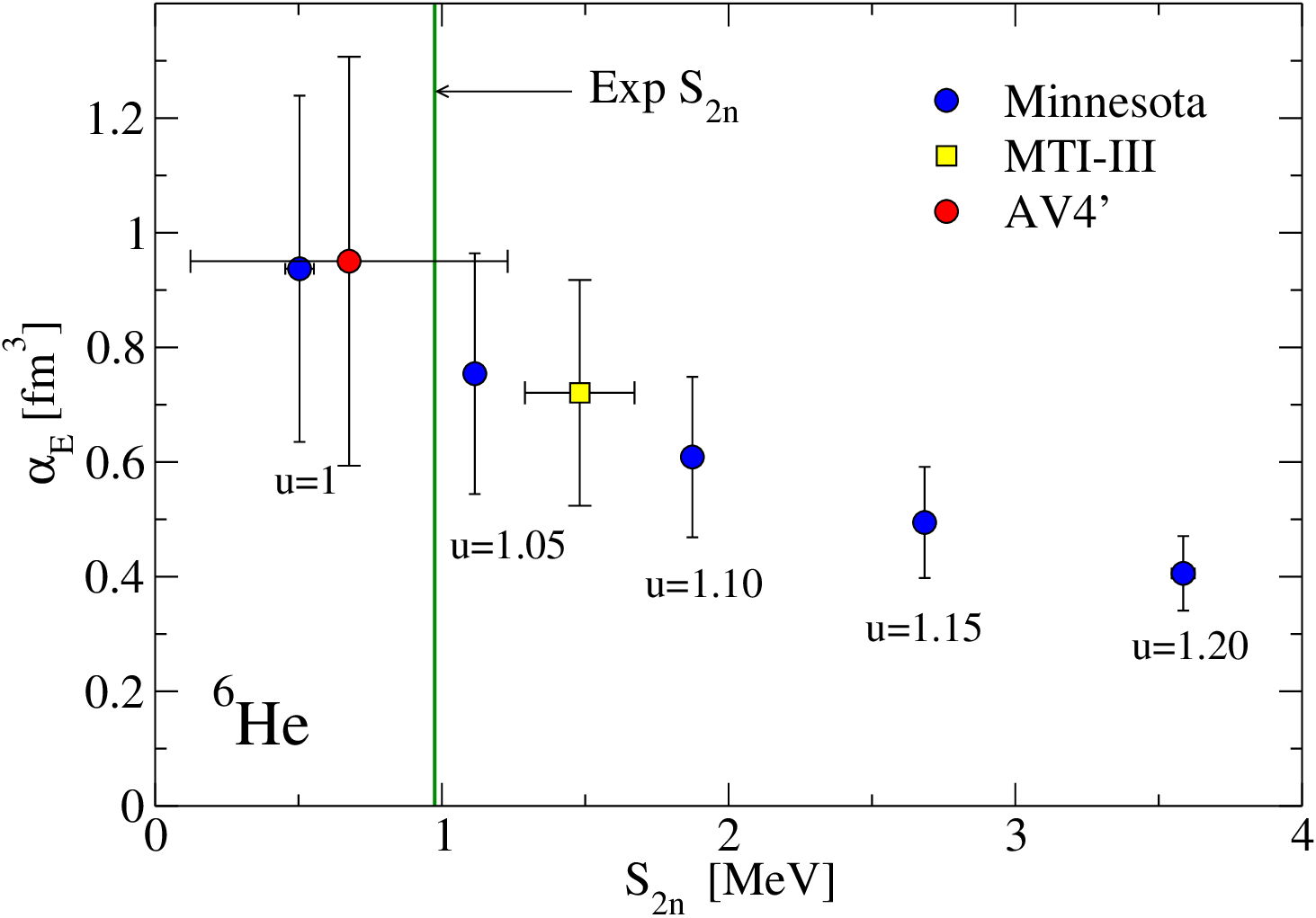}
\caption{(Color online) Correlation between $\alpha_E$ and
 $S_{2n}$ in $^6$He obtained with
the Minnesota potential varying the parameter $u$. The MTI-III and AV4' results
are also shown. 
} 
\label{fig_correlation}
\end{figure}

Now we would like to investigate the
dependence of the polarizability on the two-neutron separation energy. This can be achieved for example
 by plotting $\alpha_E$ versus $S_{2n}$ for the different values of 
the parameter $u$ in the  Minnesota potential.
In Fig.~\ref{fig_correlation}, we can see that we find a correlation between $\alpha_E$ and  $S_{2n}$.
Calculations
have been performed with $K_{\rm max}=12$ ($K'_{\rm max}=13$). As an estimate of the theoretical error bar in the few-body method
we take  the difference between the largest possible calculation  with $K_{\rm max}=12$ and the
$K_{\rm max}=8$ result. We also present the data for the MTI-III and AV4'
potential (as used in \cite{He6Sonia2}) for completeness. The error bars for
the polarizability increase as the separation energy gets smaller. This is a
reflection of the slower convergence observed in Fig.~\ref{fig_alphaE}. 
For the Minnesota potential $S_{2n}$ has a negligible error, hardly visible 
in Fig.~\ref{fig_correlation}.
For the MTI-III and AV4' potentials the error in $S_{2n}$ is large because these interaction models are not as soft as the Minnesota force.

In Ref.~\cite{Stetcu} it was argued that the polarizability should roughly scale like the inverse square of the binding energy of a nucleus. For a halo system, like $^6$He, the relevant scale parameter 
is the separation energy, rather than the binding energy. The $\alpha_E-S_{2n}$ dependence empirically observed in Fig.~\ref{fig_correlation} is compatible with such a behavior.

In order to reproduce the polarizability of a halo nucleus, it is expected that the halo structure, thus $S_{2n}$, should be correctly modeled, even if the absolute binding of $^4$He and $^6$He are not reproduced. Thus, one can estimate the value of $\alpha_E$ by choosing  $S_{2n}$ to be around the experimental value  and then calculate the corresponding polarizability. A value of $u$ that gives $S_{2n}$ close to experiment is $u=1.05$, where the convergence of $\alpha_E$ is slower than for larger values of $u$. 
From a closer look to Fig.~\ref{fig_alphaE} and Fig.~\ref{fig_S2n} we can see that  also for $u=1.20$
the polarizability $\alpha_E$ is still increasing when $K_{max}$ becomes larger, even though
the separation energy is converged. 
This means that the convergence of the polarizability is not only influenced by $S_{2n}$.
Other observable that is naturally related to the polarizability in the unretarded dipole approximation is the radius operator. 

\begin{figure}
\includegraphics[scale=0.35,clip=]{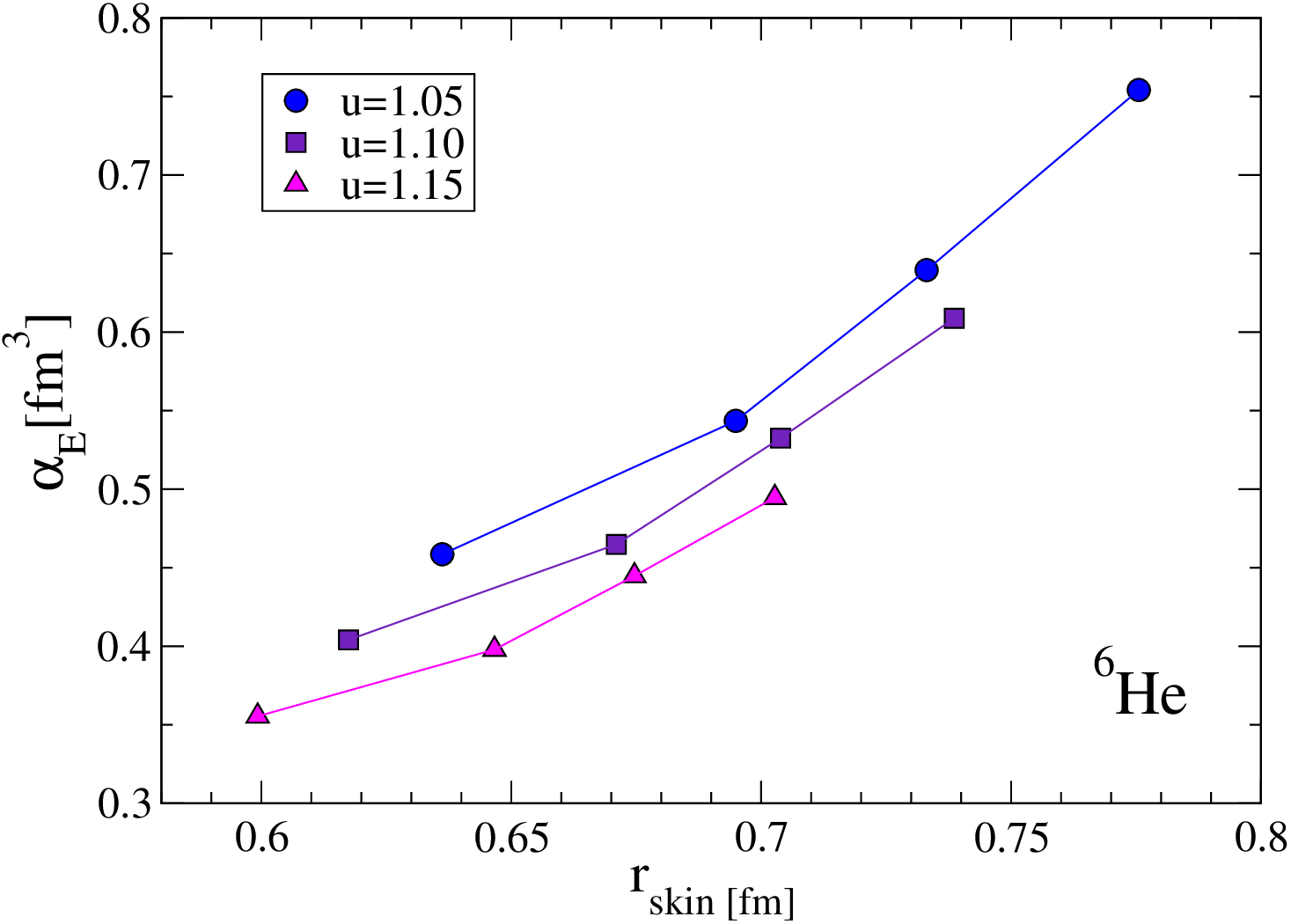}
\caption{(Color online) Correlation between the nuclear electric polarizability of $^6$He and the skin radius with the Minnesota potential with different $u$. The four points for each $u$ value correspond to calculations with $K_{\rm max}=6,8,10$ and 12, moving from the left to the right.
} 
\label{fig_alpha_skin}
\end{figure}

\begin{figure}
\includegraphics[scale=0.35,clip=]{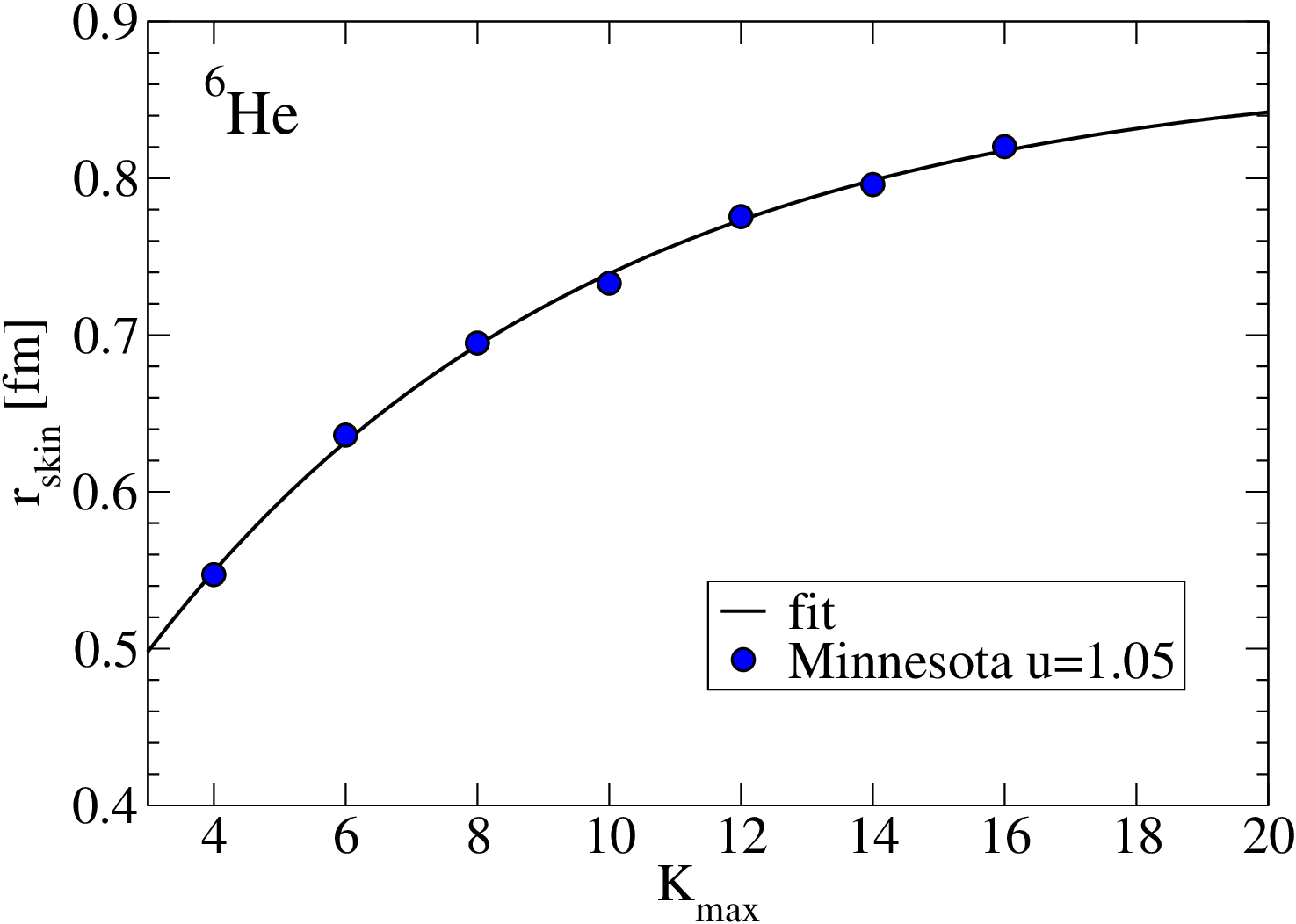}
\caption{(Color online) Neutron skin radius $r_{\rm skin}$  of $^6$He with the Minnesota potential and $u=1.05$, as a function of the grand-angular momentum quantum number $K_{\rm max}$. 
The curve is a fit to the calculated points, used to extrapolate to infinite model space.
} 
\label{rskin_fit}
\end{figure}

In  recent papers \cite{Pb_theory, Shlomo} the correlation between the
polarizability and the neutron skin of the $^{208}$Pb nucleus was studied
within the nuclear density functional theory framework. 
In the following, we will  investigate the same correlation for $^6$He, even though $^6$He is a  different system. For halo nuclei, one refers to the halo radius, rather than the skin radius, but clearly the observable
\begin{equation}
r_{\rm skin}=r_n -r_p \,,
\end{equation} 
where $r_n$ and $r_p$ are the mean point-neutron and point-proton radii, can be uniquely defined.
In our recent work~\cite{sonia_radii}, such observables have been calculated  for $^6$He from realistic two-body potentials in the EIHH method. Here, instead, we use the same semi-realistic interaction as for the $\alpha_E$ calculations.
 In Fig.~\ref{fig_alpha_skin}, we show a plot of $\alpha_E$ versus $r_{\rm skin}$
for different model spaces and for three different values of $u$ in the Minnesota potential.
The four points for each $u$ value correspond to calculations with
$K_{\rm max}=6,8,10$ and $12$, from the lowest to the largest value of $\alpha_E$,
respectively. 
For $K_{\rm max}\geq 8$ we clearly see a linear dependence between
  $\alpha_E$ and $r_{\rm skin}$ for all the three $u$ values 
where the coefficients depend on the separation energy as
\begin{equation}
\label{linear}
\alpha_E=a(S_{2n})+b(S_{2n})r_{\rm skin}\,.
\end{equation}
Because  $S_{2n}$ is converged and because of the linear dependence displayed
in Fig.~\ref{fig_alpha_skin} we deduce that the
calculation of $\alpha_E$ is 
not fully converged because the radii, and especially $r_n$, are not fully
converged. The calculation of a radius of the ground state does not require an
expansion on the dipole excited states as in Eq.~(\ref{start}) and as such is
less computation demanding and can be performed for larger model spaces, where
radii are better converged. 
Thus, the approach we take to estimate $\alpha_E$ from our calculations is to 
fit the coefficients $a(S_{2n})$ and $b(S_{2n})$ from the
$\alpha_E$ results in the available model spaces and, assuming that this
physical linear dependence  will be unchanged in larger model spaces, we will
use the coefficients to obtain $\alpha_E$ from a bound state calculation of
$r_{\rm skin}$. 
Starting with $u=1.05$, so that  $S_{2n}$ is close to
  experiment, we fit the parameters $a$ and $b$ to the 
  results of our calculations using the available values of $K_{\rm max}\geq 6$.
  We test this procedure on the
  available model space by varying the largest $K_{\rm max}$.  
  For a model space with largest $K_{\rm max}=10$, we  obtain $a=-0.7\pm0.2$~fm$^3$
and $b=1.83\pm0.3$~fm$^2$ fitting to
three points, $K_{\rm max}=6,8,10$. Using these values and the value $r_{\rm
  skin}=$0.776 fm, calculated in the  
next largest model space $K_{\rm max}=12$, our linear ansatz of Eq.~(\ref{linear})
 yields $\alpha_E=0.7\pm0.3$ fm$^3$.  
 The calculated value of $\alpha_E$ from the hyperspherical harmonics
 expansion up to $K_{\rm max}=12$ is  
 0.754 fm$^3$, which is within our estimated error band.
Now we will repeat this procedure utilizing our best three values
$K_{\rm max}=8,10,12$ (we omitted the $K_{\rm max}=6$ point as it does not fall
in line with the other points). The resulting values are 
 $a=-1.27\pm0.04$ fm$^3$ and $b=2.62\pm0.05$ fm$^2$. 
We then calculate $r_{\rm skin}$ up to the largest grand-angular momentum value
  accessible with our computational facility,
$K_{\rm max}=16$. 
Using the corresponding $r_{\rm skin}=$0.82 fm and propagating the
fit errors on $a$ and $b$ in the linear ansatz, we obtain $\alpha_E=0.88\pm0.06$
fm$^3$. Concerning the skin radius, 
one can clearly see from
Fig.~\ref{rskin_fit}, that convergence is approached. 
Extrapolating these points with an exponential {\it ansatz} of the form~ $r_{\rm skin}(K_{\rm max})=r_{\rm skin}({\infty})-c e^{-\kappa K_{\rm max}}$
we get $r_{\rm skin}(\infty)=0.87(5)$ fm. As an error estimate
we take the difference between $r_{\rm skin}(K_{\rm max}=16)$ and  $r_{\rm skin}(K_{\rm max}=12)$.
The theoretical value is somehow larger than the experimental data. 
In fact, combining different measurements of the matter radius \cite{Tan92,Alk97,Kis05} with the most recent evaluation  of the proton radius \cite{Maxime},  one can infer $r_n$ and consequently the skin radius, which is found to be  0.52$\le r_{\rm skin}^{\rm exp}\le$0.62 fm.
The variation on $r_{\rm skin}^{\rm exp}$ is fairly large, due to the large
uncertainty in the matter radius determination from ion scattering. 

Because the extrapolated skin radius is our best estimate of this observable, we use this value in
Eq.~(\ref{linear}) and propagate its error 
estimating the polarizability, considering it independent from the fit errors on $a$ and $b$. 
Finally, our estimate of the theoretical nuclear electric polarizability of $^6$He is  
$\alpha_E=1.00(14)$ fm$^3$. This value is consistent  with
what we obtained without extrapolating the radius, showing that the error bars are based on conservative estimates.
 If we were to use Eq.~(\ref{linear}) with the
experimental values of the skin radius, one would obtain a nuclear electric polarizability of  $\alpha_E=0.08-0.34$ fm$^3$, which is even smaller than the estimate based solely on theory.

\section{Comparison with experiment}

\begin{figure}
\includegraphics[scale=0.35,clip=]{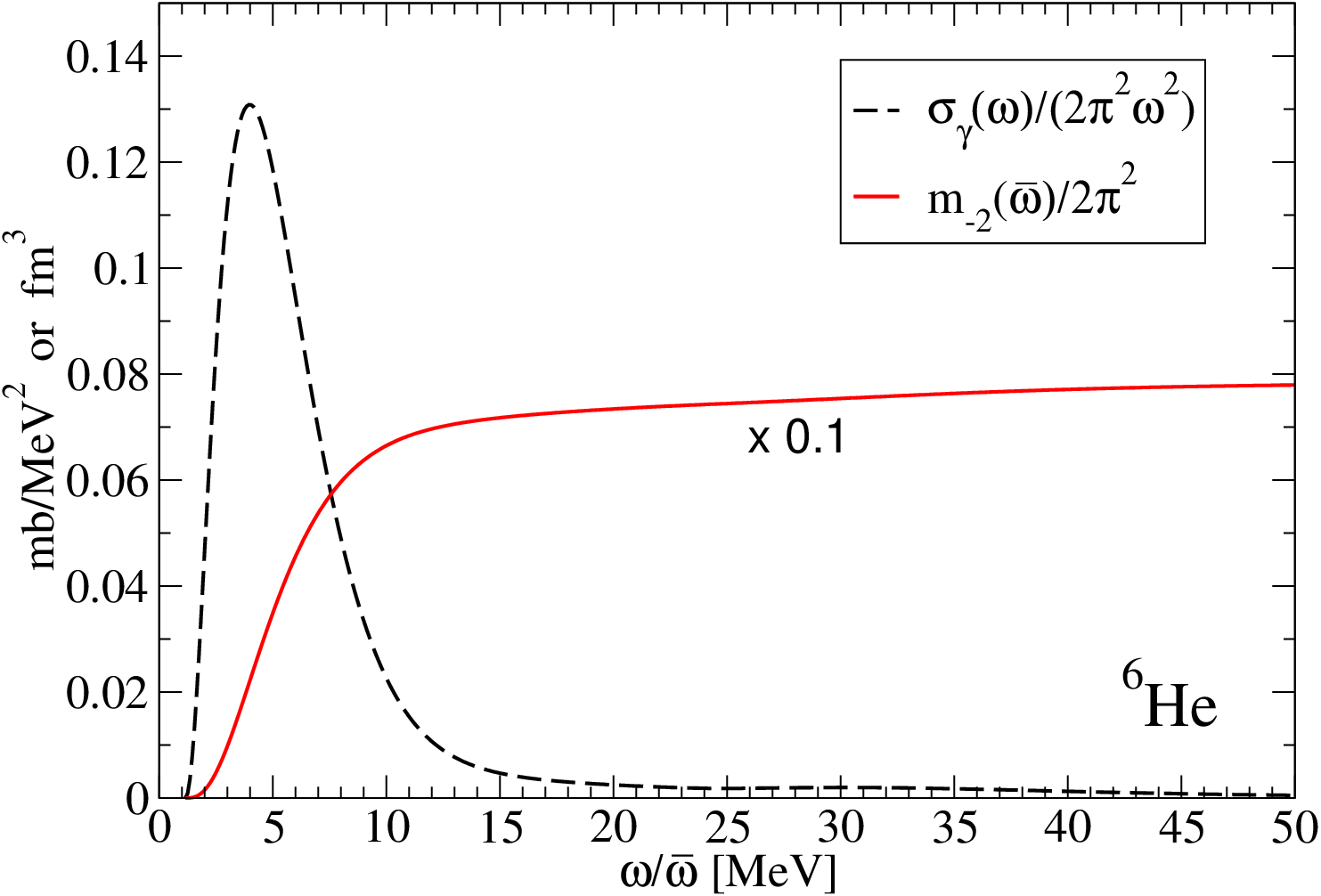}
\caption{(Color online) Double inverse energy weighted cross section as a function of the energy $\omega$ and sum rule $m_{-2}(\bar{\omega})/2\pi^2$ as a function of $\bar{\omega}$ for $^6$He with the Minnesota potential and $u=1.05$.
} 
\label{fig_integral}
\end{figure}

\vspace{0.5cm}
\begin{figure}
\includegraphics[scale=0.35,clip=]{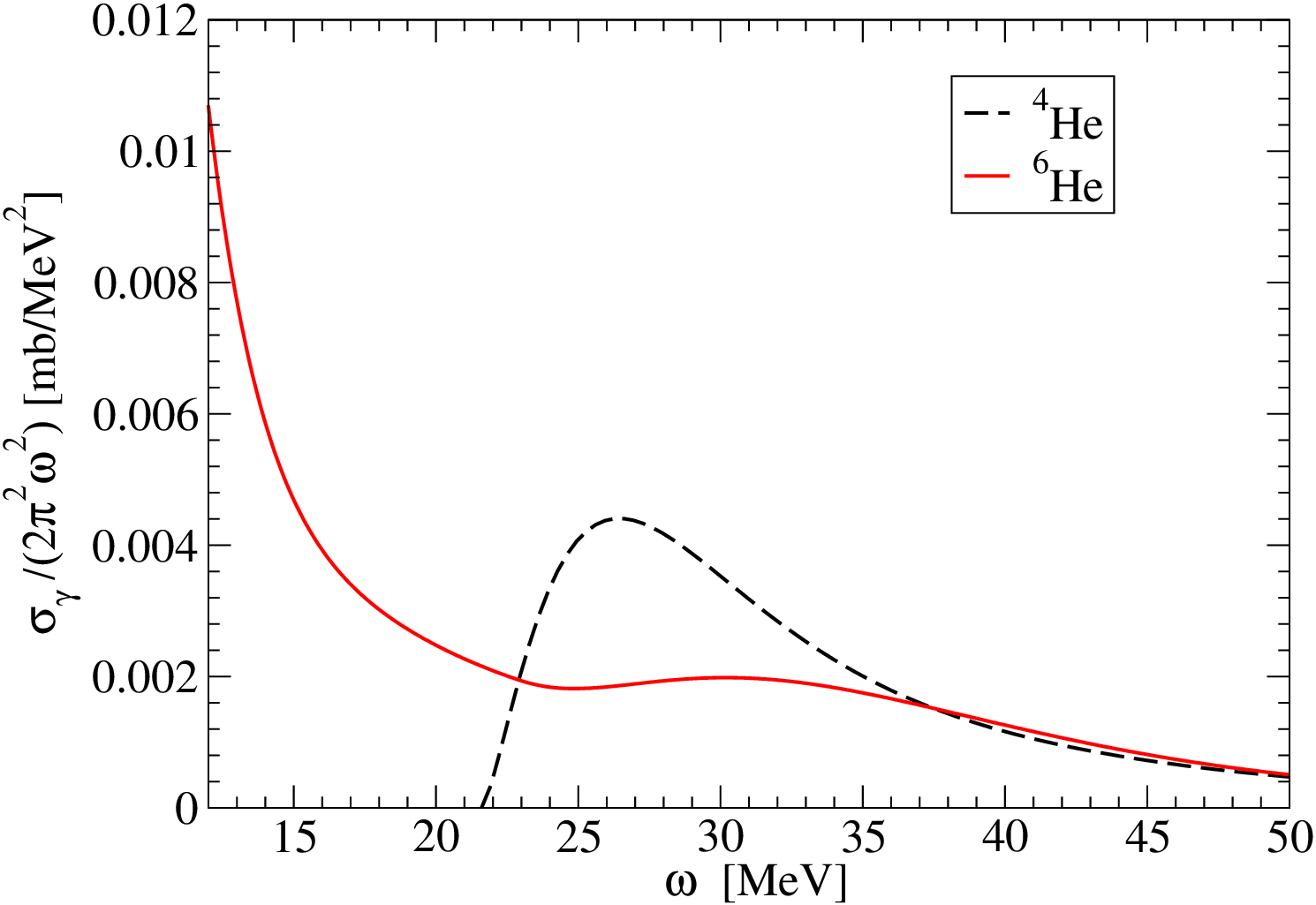}
\caption{(Color online) Double inverse energy weighted cross section as a function of the energy $\omega$ for $^4$He and $^6$He with the Minnesota potential and $u=1.05$. 
} 
\label{fig_comp}
\end{figure}

In Ref.~\cite{Moro} an evaluation of the experimental number for the
polarizability of $^6$He was presented, leading to  $\alpha^{\rm exp}_E=1.99(40)$ fm$^3$.
This was obtained from the inverse energy weighted integral of experimental and  theoretical $B(E1)$ response functions for $^6$He. The experimental distribution was measured by Aumann {\it et al.} 
\cite{Aumann} from the Coulomb breakup of $^6$He on Lead and Carbon at 240 MeV/$u$ up to 8 MeV above threshold. In order to obtain the polarizability, data were extrapolated up to 12.3 MeV, where the threshold for the breakup into two tritons opens up. The theoretical curve was taken from a calculation of the dipole transition to the $1^-$ continuum  \cite{Thompson} 
 in a three-body model with phenomenological $n-n$  and $n-\alpha$ interactions plus an effective three-body force, also calculated up to the two tritons threshold.
The estimate was done in two steps: ({\it i}) an average of the experimental data and theoretical curve was taken up to 12.3 MeV; ({\it ii})
to account for the higher energies, the polarizability of $^4$He was added. The latter one basically comes from integrating the photo-dissociation data  from Arkatov {\it et al.} \cite{Arkatov}.
Because we can access the full response functions using the LIT method at any energy below the pion production threshold, we can verify these two approximations. First, it is interesting to study 
 the convergence of $\alpha_E$ calculated as a sum rule of the response, see Eq.~(\ref{moments}) and (\ref{alphaEPSR}), to investigate the validity of ({\it i}).
In  Fig.~\ref{fig_integral},  we present both the integrand function $\sigma_{\gamma}(\omega)/2\pi^2\omega^2$ versus $\omega$ and the convergence of the integral $m_{-2}(\bar{\omega})/2\pi^2$ versus $\bar{\omega}$.
At $\bar{\omega}=8$ MeV the sum rule is exhausted only up to 75$\%$. Thus only 75$\%$ of the $\alpha_E^{\rm exp}$ is based solely on experimental data.
At $\bar{\omega}=$12.3 MeV, where the two $^3$H channel opens up  the sum rule is exhausted up to
90$\%$. We observe that one needs to integrate up to $40$ MeV of energy to have the sum rule exhausted at the $98\%$ level. 

To verify the approximation ({\it ii}) we can compare the 
integrand function $\sigma_{\gamma}(\omega)/2\pi^2\omega^2$ for $^6$He and $^4$He at energies beyond 12.3 MeV. In Fig.~(\ref{fig_comp}) we observe that the two curves agree with each other for $\omega>35$ MeV, where the sum rule is almost exhausted. In the region beyond the $^4$He disintegration threshold and below about $35$ MeV one would over estimate the sum rule integrating the $^4$He curve, because  one gets 0.044 fm$^3$, to be compared to the 0.027 fm$^3$ obtained when correctly integrating the $^6$He curve. On the other hand, neglecting the part of the cross section for $\omega>12.3$ MeV and below the $^4$He disintegration threshold one underestimates the sum rule. The  contribution of this portion is  0.037 fm$^3$, about 5$\%$ of the sum rule. These two effects almost cancel out so that, the approximation ({\it ii}) does not lead to a big error.

We think that the main reason of the disagreement between the estimate from Ref.~\cite{Moro} and our
calculations comes from the difference in the low-energy part of the
response. In our previous work \cite{He6Sonia1, He6Sonia2} we have shown that
our calculations with semi-realistic potentials underestimates the data from
Aumann {\it et al.}~\cite{Aumann}. Thus, what we observe for the polarizability is consistent
with this fact. We would like to point out that (i) nuclear corrections might
affect the results in the ion scattering experiment of \cite{Aumann} and that (ii) as
discussed earlier, similar experiments for $^3$He 
lead to large discrepancy with photo-dissociation results.
Nevertheless, measuring $\alpha_E$ from the dipole response
function it would be desirable to have data up to higher energies than 12.3
MeV. 
Additional or alternative measurements of $\alpha_E$ would help to better constrain this observable.

\section{Conclusions}
We summarize our results as follows. We have carried out an estimate 
 of the nuclear polarizability of $^6$He based on the hyperspherical harmonics expansion
with simple semi-realistic potentials.
Our calculations clearly reproduce the fact that the polarizability of the halo nucleus of $^6$He is much larger than that of the tightly bound $^4$He.
For $^4$He the semi-realistic Minnesota potentials lead to a value of the polarizability which is consistent with realistic calculations and is about 15$\%$ smaller than the average value in the experimental band. Nevertheless, a large disagreement is found for $^6$He. 
In order to estimate $\alpha_E$
we have chosen a potential that reproduces the separation energy and then we investigated the correlation  of the polarizability with the skin radius. 
Our final result is 
$\alpha_E=1.00(14)$ fm$^3$, 
which is about a factor of 2
smaller than the estimates from experimental data. This points towards a disagreement of microscopic theory and
experiments. To shed light on this, it would be nice to have more data or
alternative measurements of $\alpha_E$. Concerning the theoretical
calculations, it is desirable 
to extend these results to realistic potentials including also three-body forces. We leave this subject to a future work.
\vspace{0.5cm}

\section{Acknowledgment}

The work of  R.~Goerke and S.~Bacca
was supported in part by the
Natural Sciences and Engineering Research Council (NSERC) and in part by the
National Research Council of Canada.
The work of N. Barnea was supported by the ISRAEL SCIENCE FOUNDATION
(grant no.~954/09).
Numerical calculations were  performed at TRIUMF.

\end{document}